\documentclass{elsart1p}
\newcommand{\be}{\begin{equation}}
\newcommand{\ee}{\end{equation}}
\newcommand{\xx}{{\mathbf x}}
\newcommand{\kk}{{\mathbf k}}
\newcommand{\vv}{{\mathbf v}}
\newcommand{\eq}[1]{(\ref{#1})}

\newcommand{\mume}{\mu \rm m}
\newcommand{\im}{\textrm{Im}}

\newcommand{\rr}{{\bf r}}
\newcommand{\prl}[3]{Phys. Rev. Lett. {\bf #1}, #2 (#3)}

\usepackage[active]{srcltx}

\usepackage{graphicx}

\usepackage{amssymb}

\begin{document}

\begin{frontmatter}


\title{Excitations and superfluidity in non-equilibrium Bose-Einstein condensates
of exciton-polaritons}
\author{Michiel Wouters}
 \ead{michiel.wouters@ua.ac.be}
 \address{TFVS, Universiteit Antwerpen, Groenenborgerlaan 171, 2020 Antwerpen, Belgium}


\author{Iacopo Carusotto}

\address{BEC-CNR-INFM and Dipartimento di Fisica, Universit\`a di Trento, I-38050 Povo, Italy}

\begin{abstract}
We present a generic model for the description of non-equilibrium
Bose-Einstein condensates, suited for the modelling of non-resonantly
pumped polariton condensates in a semiconductor microcavity. The
excitation spectrum and scattering of the non-equilibrium
condensate with a defect are discussed. 
\end{abstract}

\end{frontmatter}


\section{Model and elementary excitation spectrum}

Due to the finite polariton life time, the nature of polariton condensates~\cite{nonres-expt} is fundamentally different
from their superfluid $^4$He and atomic Bose-Einstein condensate counterparts: the polariton condensate has to be constantly replenished and arises as a dynamical equilibrium between pumping and decay. The theoretical modelling of such systems in which interactions, coherence, pumping and decay are equally important poses a challenge that was taken up only recently \cite{littlewood,sarchi,generic}. We will present here our mean field model of such condensates~\cite{generic} and use it to study the excitation spectrum of a homogeneous non-equilibrium condensate and the problem of a small defect moving through it.

Our model does not include any details on the specific relaxation mechanisms
of high-energy polaritons into the condensate, but is only based on some
general assumptions: i) a single state of lower polaritons is macroscopically occupied so that it can be described by a classical field; ii) The momentum space can be devided in two parts: one part at small momenta where the coherence is important and one part at large momenta, where it is negligble. The polaritons in the high-momentum states act as a reservoir that replenishes the condensate. Under typical excitation conditions, the wave vector scale to separate both systems should be chosen of the order of a few $\mume^{-1}$; iii) The state of the reservoir is fully determined by its spatial polariton density $n_R(x)$. This last asumption requires that the reservoir polariton momentum distribution reaches some stationary state in momentum space. Under these assumptions, the condensate dynamics is to a first approximation described by a generalized Gross-Pitaevskii equation including loss and amplification terms
\begin{equation}
i\frac{\partial \psi}{\partial t}=\left\{ -\frac{\hbar \nabla ^{2}}{2m}+
\frac{i}{2}\big[R(n_R)-\gamma\big]+g\,|\psi|^{2}+2\tilde{g}\,n_{R}\right\}
\psi ,  \label{eq:GP}
\end{equation}
where $m$ is the lower polariton mass, $\gamma$ its decay rate and
$g$ is the strength of the polariton-polariton interaction within the
condensate. The stimulated scattering of reservoir polaritons into the
condensate is modeled by the term $R(n_R)$ and the mean field
interaction experienced by the condensate polaritons due to elastic
collisions with the reservoir polaritons is given by $2\tilde g n_R$.
The description \eq{eq:GP} of the polariton condensate in terms of a deterministic classical field requires its density and phase fluctuations to be small. This regime is reached for pump powers well above the condensation threshold. The equation for the condensate dynamics is coupled to a diffusion equation for the reservoir polaritons
\begin{equation}
\frac{\partial n_R}{\partial t}=P-\gamma _{R}\,n_{R}-R\left(
n_{R}\right) \left\vert \psi \left( x\right) \right\vert ^{2}+
D\nabla ^{2}n_{R},  \label{eq:rate_B}
\end{equation}
where $P$ is the pump rate due to external laser, $\gamma_R$ is the
reservoir damping rate and $D$ its diffusion constant.

The stationary state and the elementary excitation spectrum of Eqns. \eq{eq:GP} and \eq{eq:rate_B} are discussed in Ref.~\cite{generic}. In case the reservoir damping rate $\gamma_R$ is much larger than the condensate polariton damping rate $\gamma$, the condensate excitation spectrum is of the form
\begin{equation}
\omega_\pm(k) =-\frac{i\Gamma}{2}\pm \sqrt{\omega _{Bog}(k)^2 -
\frac{\Gamma^{2}} {4}},  \label{spectr}
\end{equation}
Here, $\omega_{Bog} = [(k^2/2m+2 \mu)(k^2/2m)]^{1/2}$ is the usual Bogoliubov dispersion of dilute Bose gases at equilibrium. The non-equilibrium nature of the system is quantified by the effective relaxation rate $\Gamma=\zeta \gamma$, where $\zeta$ depends on the pumping rate and on the functional form of $R(n_R)$\cite{generic}. The most important differences with the excitation spectrum of equilibrium condensates is the non vanishing imaginary part of $\omega(k)$ for all $k\neq 0$ and the flatness of its real part for small $k$. The `+'-branch is diffusive for small wave vectors. A similar excitation spectrum was found in Ref.  \cite{littlewood} for a specific model of nonequilibrium condensation, within a completely different approach, indicating that the form \eq{spectr} is a general result for nonequilibrium condensates.

Our model is also straightforwardly applied to the Josephson
oscillations between two condensates connected by quantum mechanical
tunneling. The frequency of the density oscillations between the two wells 
is given by Eq.\eq{spectr} with $\omega_{Bog}$ replaced by the equilibrium Josephson frequency~\cite{milburn}.

\section{Flow past a defect}

One of the benchmark properties of condensed Bose systems is
superfluidity. As a first step in the study of superfluidity in
non-equilibrium systems, we will discuss the scattering of a moving
condensate on a defect \cite{iac-sup}.  
Experimentally, the condensate could be
accelerated by applying an external force to it, e.g. by making a
sample with a steep wedge in the cavity thickness or using surface
acoustic waves to accelerate the polaritons \cite{lima}. Defects are
naturally present in the form of disorder, but can also be
deliberately created by structuring the cavity mirrors \cite{tao}.

\begin{figure}[htbp]
\begin{center}
\includegraphics[width=0.5\columnwidth,angle=0,clip]{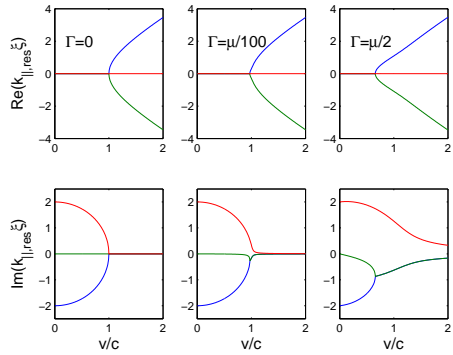} 
\end{center}
\caption{The real (upper panels) and corresponding imaginary (lower
  panels) part of the spatial frequencies (in units of the inverse of the healing length $\xi=\sqrt{\hbar/m\mu}$) that are excited by a
  stationary defect in a condensate moving at a speed $v$ 
  for several values of $\Gamma/\mu$.}
\label{fig:re-imk3}
\end{figure}

The perturbation on top of a condensate moving with velocity $\vv$ due to a
small defect potential $V_{def}=g\delta(\rr)$ at rest can be studied
in perturbation theory \cite{iac-sup}. The change in condensate wave
function is in momentum space given by
\begin{equation}
[\delta \psi(\kk), \delta \psi^*(2\kk_0-\kk), \delta n_r]^T=- \mathcal L_\vv(\kk)^{-1} [g \psi_0(0),\; -g \psi_0^*(0), \; 0]^T,
\end{equation}
where $\mathcal L_\vv(\kk)$ the matrix from linearized motion
equations \eq{eq:GP},\eq{eq:rate_B} around a steady state
$n_R(\xx)=n_0$, $\psi(\xx)=\psi_0e^{i\kk_0\cdot \xx}$, where $\kk_0=m \vv$. Much about the
response of the flowing condensate is learned by studying the poles $\kk_{res}$ of $L_\vv(\kk)^{-1}$ in the complex $\kk$-plane, i.e.
\begin{equation}
\omega_{\vv \pm}(\kk_{res})=\omega_\pm(\kk_{res}-\kk_0)-\vv\cdot(\kk_{res}-\kk_0)=0,
\label{eq:bog_gal}
\end{equation}
We restrict now our attention to the perturbation of the wave function in the direction of the defect velocity. The main contribution comes from the Fourier component with $\kk \parallel \vv$. The modulus of the resonant spatial frequency $k_{\parallel res}$ is plotted in Fig.\ref{fig:re-imk3}. The panels show from left to right an equilibrium condensate ($\Gamma=0$), a slightly non-equilibrium condensate ($\Gamma/\mu=0.01$) and a condensate where interaction effects and losses are comparable ($\Gamma=\mu/2$). Unlike for the temporal frequencies from Eq. \eq{spectr}, the imaginary part of the spatial frequencies should not be negative, but the spatial frequencies with a negative (positive) imaginary part correspond to positions on the left (right) hand side of the defect.

Let us start discussing the resonant spatial frequencies of the equilibrium condensate. The phase fluctuations, that give a pole at zero spatial frequency, do not couple to the external potential.  The two other branches of $k_{\parallel res}$ are purely imaginary for velocities smaller than the speed of sound, implying that the condensate perturbation is spatially damped. The excitations are only virtually excited by the defect and no condensate momentum is dissipated. At $v=c$, there is a bifurcation, where a pair of conjugate imaginary roots turn real, the excitations go `on shell', are radiated and dissipate the condensate momentum.

In the case of non-equilibrium condensates (central and right panels of Fig.  \ref{fig:re-imk3}), the bifurcation point is shifted to velocities $v<c$ and $\im(k_{\parallel,res})<0$.  No spatial frequencies with zero imaginary part appear anymore. Therefore, in contrast to the equilibrium condensates, no sharp transition in the condensate perturbation as a function of its velocicy is expected.

\begin{figure}[htbp]
\begin{center}
\includegraphics[width=0.8\columnwidth,angle=0,clip]{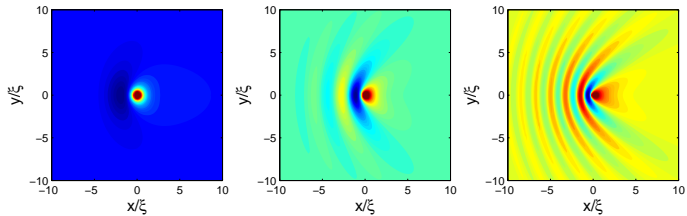} 
\end{center}
\caption{The perturbation of the condensate moving condensate due to a
  small defect at $x=0$. Condensate velocities: $v/c=0.7$ (left
  panel), $v/c=1.3$ (center panel) and $v/c=2$ (right
  panel). Effective damping rate $\Gamma=\mu/2$. }
\label{fig:superf_real}
\end{figure}

This is confirmed by the condensate perturbation in real space, shown in Fig. \ref{fig:superf_real} for several velocities $v$. The condensate perturbation depends sensitively on the velocity $v$, but no sharp transition occurs. It is smoothened due to the finite polariton life time.



\end{document}